\documentclass[10pt,aps,showpacs,nofootinbib,prd,twocolumn,showkeys,floatfix]{revtex4-1}

\usepackage{graphicx}
\usepackage{epstopdf}
\usepackage{epsfig}
\usepackage{pgfplots}
\pgfplotsset{compat=1.18}

\usepackage{tikz}
\usepackage[compat=1.1.0]{tikz-feynman}
\usepackage[compat=1.1.0]{tikz-feynhand}
\usetikzlibrary{arrows.meta, positioning, decorations.markings, calc, patterns}
\usepackage{amsmath}
\usepackage{amssymb}
\usepackage{mathtools}
\usepackage{slashed}
\usepackage{physics}
\usepackage{IEEEtrantools}  

\usepackage{newtxtext}
\usepackage{newtxmath}
\usepackage[normalem]{ulem}

\usepackage[svgnames]{xcolor}
\usepackage{url}
\usepackage{subfiles}  

\usepackage[colorlinks, citecolor=DarkGreen, linkcolor=DarkRed, urlcolor=DarkBlue]{hyperref}

\newcommand{\lag}{\mathcal{L}}

\newcommand{\be}{\begin{eqnarray}}
\newcommand{\ee}{\end{eqnarray}}
\begin{document}

\title{Chiral first order phase transition at finite baryon density and zero temperature from self-consistent pole masses in the linear sigma model with quarks}

\author{Alejandro Ayala$^1$}
\author{Bruno El-Bennich$^{2,3}$}
\author{Ricardo L. S. Farias$^{4,5}$}
\author{Luis A. Hern\'andez$^6$}
\author{Bruno S. Lopes$^{4,5}$}
\author{Luis Carlos Parra Lara$^1$}
\email[Corresponding author: ]{luis.parra@correo.nucleares.unam.mx}
\author{Renato Zamora$^{7,8}$}

\affiliation{
$^1$Instituto de Ciencias Nucleares, Universidad Nacional Aut\'onoma de M\'exico, Apartado
Postal 70-543, CdMx 04510, Mexico}
\affiliation{$^2$Departamento de F\'isica, Universidade Federal de S\~ao Paulo, Rua S\~ao Nicolau 210, 09913-030 Diadema, S\~ao Paulo, Brazil}
\affiliation{$^3$Instituto de F\'\i sica Te\'orica, Universidade Estadual Paulista, Rua Dr. Bento Teobaldo Ferraz 271 - Bloco II, 01140-070 S\~ao Paulo, S\~ao Paulo, Brazil}
\affiliation{$^4$Departamento de F\'\i sica, Universidade Federal de Santa Maria, RS 97105-900, Rio Grande do Sul, Brazil}
\affiliation{$^5$Center for Nuclear Research, Department of Physics, Kent State University, Kent, OH 44242, USA}
\affiliation{$^6$Departamento de F\'\i sica, Universidad Aut\'onoma Metropolitana-Iztapalapa,
Avenida San Rafael Atlixco 186, Ciudad de M\'exico 09340, Mexico}
\affiliation{$^7$Centro de Investigaci\'on y Desarrollo en Ciencias Aeroespaciales (CIDCA), Academia Politécnica Aeronáutica, Fuerza A\'erea de Chile, Casilla 8020744, Santiago, Chile.}
\affiliation{$^8$Facultad de Ingenier\'\i a, Arquitectura y Dise\~no, Universidad San Sebasti\'an, Santiago, Chile}

\begin{abstract}

We use the two-flavor Linear Sigma Model with quarks as an effective description of QCD to investigate the nature of the chiral phase transition at finite baryon chemical potential and zero temperature. We work at one-loop order to set up and solve the system of self-consistent coupled equations for the particle pole masses. The chemical potential-dependent value of the chiral order parameter is obtained by minimizing the one-loop effective potential. This treatment goes beyond the conventional ring-diagram approximation and provides a description valid for arbitrary values of the chemical potential. We find that the phase transition is of first order, and occurs when the quark chemical potential reaches the value of the vacuum quark mass for the chosen set of parameters. The first order nature of the transition is signaled by the discontinuous behavior of the chiral condensate, the masses and the couplings. The thermodynamics of the system is readily implemented and in particular, we find that the square of the speed of sound exhibits a discontinuity at the phase transition and then smoothly approaches the conformal limit from below.

\end{abstract}


\maketitle

\section{Introduction}

The phase structure of Quantum Chromodynamics (QCD) at finite temperature ($T$) and baryon density ($\mu_B$) reflects the interplay between deconfinement and chiral symmetry restoration. At low baryon densities, this transition proceeds as a smooth crossover, as established by both experimental and theoretical studies. At higher densities, however, the crossover is expected to terminate at a critical end point (CEP) or be replaced by more complex phases, such as those with spatial modulations, potentially originating the onset of qualitatively new regimes in the QCD phase diagram~\cite{Fukushima:2010bq,Pisarski:2021qof,Rennecke:2023xhc,Arslandok:2023utm,Luo:2022mtp,Stephanov:1999zu,Stephanov:2008qz,Luo:2017faz,Mroczek:2020rpm}. In this context, the properties of strongly interacting matter under extreme conditions of temperature, density, and also in the presence of strong electromagnetic fields are currently under intense scrutiny~\cite{Fu:2026qnl,Fu:2019hdw,Gutierrez:2013sta,Brandt:2024dle,Ayala:2006gf,Braun:2009gm,Fukushima:2013rx,Andersen:2014xxa,Farias:2014eca,Miransky:2015ava,Adhikari:2019mdk,Endrodi:2024cqn,Adhikari:2024bfa,Ding:2025nyh,Hamada:2026uec,Schmidt:2025ppy,Ayala:2024sqm,Lopes:2021tro,Lopes:2025rvn,Ayala:2023mms,Ayala:2023cnt,Pasqualotto:2023hho,Hernandez:2025wye}. 

A driving force in this field is the experimental search for the CEP, for which the STAR BES-I and BES-II programs have provided a wealth of information on net baryon number fluctuations~\cite{STAR:2021rls,STAR:2022vlo,STAR:2021fge,STAR:2022etb,Zhang:2026dny,Pandav:2024llq}. In the near future, the Multi-Purpose Detector (MPD)~\cite{MPD:2022qhn,MPD:2025jzd}, at the Nuclotron-based Ion Collider Facility (NICA), will provide additional data, precisely in the region of the phase diagram where the CEP is thought to be located, which corresponds to the NICA energy range.

If a CEP exists, the transition line between broken and restored chiral symmetry for large baryon densities turns out to be first order. Therefore, exploring the region of large $\mu_B$ and low $T$ becomes important in the search for signatures of this kind of transition. Unfortunately, Lattice QCD cannot perform this search deep into the phase diagram due to the severe sign problem~\cite{Troyer:2004ge,Nagata:2021ugx,Schmidt:2025ftp}. To circumvent this situation, several theoretical tools have been explored, such as the analysis of the Lee-Yang zeroes in the complex chemical potential plane~\cite{Basar:2023nkp,Wan:2024xeu}, Schwinger-Dyson equations~\cite{Fischer:2018sdj,Bernhardt:2021iql}, the functional renormalization group~\cite{Tripolt:2013jra,Fu:2022gou}, Ising mapping to the QCD CEP~\cite{Parotto:2018pwx,Kahangirwe:2024cny}, holographic methods~\cite{Critelli:2017oub,Grefa:2021qvt}, and effective models~\cite{Farias:2005cr,Fu:2023lcm,Ayala:2019skg,Ayala:2018maf,Duarte:2018kfd,Lopes:2022efy,Nunes:2024hzy,Farias:2025vss}. In particular, the two-flavor Linear Sigma Model with quarks (LSMq) has proven to be an excellent analytical tool due to its trackable particle content (pions, sigma, and quarks) and its renormalization properties~\cite{Scavenius:2000qd,Schaefer:2004en,Schaefer:2007pw,Herbst:2010rf,Tripolt:2017zgc}.

However, pions, being pseudo-Goldstone bosons, are light, and since in the LSMq particle masses are driven by the chiral condensate, when chiral symmetry is being restored, the square of the pion mass can become negative, signaling the need to include the plasma screening effects in the model description. This approach has been followed, including the finite $\mu_B$ and $T$ static contributions to the meson masses by accounting for the meson self-energies at ring diagram order~\cite{Ayala:2021tkm,Ayala:2021nhx,Ayala:2015hba}. In addition to the static nature of the self-energy calculations, analytic results can only be obtained in the limit of large $T$, which, for the description of the chiral transition line, is at best a marginally good approximation. Attempts to extend the range of validity of the analytical approximation for smaller values of $T$ are also limited~\cite{Ayala:2017ucc}.

It is therefore clear that if these limitations are to be overcome, the $\mu_B$ and $T$ modifications to the two-flavor LSMq particle masses need to be computed beyond the static, high temperature ring diagram approximation. In this work we start the road for the computation of these particle-mass modifications, concentrating first on the case of arbitrary finite $\mu_B$ while keeping $T=0$. We work in perturbation theory at one-loop order. Since the Feynman diagrams contributing to each particle self-energy depend on the masses of the other degrees of freedom, the calculation is intrinsically self-consistent and requires the simultaneous solution of a set of three coupled integral equations for the $\mu_B$-dependent pion, sigma, and quark masses. These masses, in turn, depend on the vacuum expectation value $v$ of the sigma field, which is the order parameter of the theory, and is obtained by minimizing the one-loop effective potential. The work is organized as follows: In Sec.~\ref{II} we write the LSMq Lagrangian and describe its properties. Working up to one-loop order, we separate the vacuum from the matter contribution to the effective potential and find the expression for the matter piece in the $T=0$ limit. In Sec.~\ref{III} we explicitly compute the self-energies at finite $\mu_B$ that modify the particle masses. In Sec.~\ref{IV} we numerically solve the equations and discuss the properties of the solutions. Finally, in Sec.~\ref{concl}, we summarize our findings and provide an outlook on the results. In what follows, we work with the quark chemical potential $\mu$ and recall that its relation with the baryon chemical potential is given by $\mu_B=3\mu$.

\section{Linear sigma model with quarks}\label{II}

We consider the linear sigma model with $N_f = 2$ quark flavors and $N_c = 3$ colors. The Lagrangian density in the chiral limit can be written as
\begin{eqnarray}
        \lag &=& \frac{1}{2}\mbox{Tr}\left[\partial_\mu\Phi^\dagger\partial^\mu\Phi\right] + \frac{a^2}{4}\mbox{Tr}\left[\Phi^\dagger\Phi\right] - \frac{\lambda}{16}\mbox{Tr}\left[\Phi^\dagger\Phi\right]^2 \nonumber\\
        &+& i \bar{\psi} \slashed{\partial} \psi- g \left( \bar{\psi}_L \Phi \psi_R + \bar{\psi}_R \Phi^\dagger \psi_L \right),
        \label{eq:1}                 
\end{eqnarray}
where the squared mass parameter $a^2>0$ allows for a spontaneous chiral symmetry breaking, and $\psi_{L,R} = [(1 \mp \gamma^5)/2] \, \psi$ are the chiral projections of the quark field. The boson self-coupling and the boson–fermion coupling, $\lambda$ and $g$, respectively, are taken to be positive. The meson field is given explicitly by $\Phi= \sigma + i \,\vec{\tau} \cdot \vec{\pi }$, where $\sigma$ is the scalar sigma field and $\vec{\pi} = (\pi_1, \pi_2, \pi_3)^T$ is the triplet of pion pseudoscalars. This field transforms in flavor space under chiral transformations as $\Phi \;\rightarrow\; U_L \,\Phi \, U_R^\dagger$, with $U_{L,R} \in \text{SU}(2)_L \times \text{SU}(2)_R
$. These transformations are parametrized by two sets of angles $\Vec{\alpha}_L$ and $\Vec{\alpha}_R$, with the corresponding unitary operators
\begin{equation}
    U_L = e^{i \Vec{\alpha}_L \cdot \Vec{\tau}/2}, \qquad 
    U_R = e^{i \Vec{\alpha}_R \cdot \Vec{\tau}/2},
    \label{2}
\end{equation}
where $\Vec{\tau} = (\tau_1, \tau_2, \tau_3)^T$ are the Pauli matrices. The fermion fields transform as
\begin{equation}    
    \psi_L \to U_L \psi_L, \qquad \psi_R \to U_R \psi_R.
    \label{3}
\end{equation}
In terms of the Dirac field $\psi = \psi_L + \psi_R$, this can be written as
\begin{equation}    
    \psi \to \exp\!\left(i \frac{\Vec{\alpha}_V +\gamma^5 \Vec{\alpha}_A}{2}\cdot \Vec{\tau} \right) \psi,
    \label{4}
\end{equation}
where $\Vec{\alpha}_V = (\Vec{\alpha}_L + \Vec{\alpha}_R)/2$ (vector transformation) and $\Vec{\alpha}_A = (\Vec{\alpha}_L - \Vec{\alpha}_R)/2$ (axial-vector transformation). Expanding in components, we obtain the transformations for $\sigma$ and $\vec{\pi}$.

\noindent\textbf{Vector transformation} ($\Vec{\alpha}_V$):
\begin{equation}    
    \sigma \to \sigma, \qquad \vec{\pi} \to \vec{\pi} + \Vec{\alpha}_V \times \vec{\pi}.
    \label{5}
\end{equation}
This is an isospin rotation, leaving $\sigma$ invariant.

\noindent\textbf{Axial-vector transformation} ($\Vec{\alpha}_A$):
\begin{equation}        
    \sigma \to \sigma + \Vec{\alpha}_A \cdot \vec{\pi}, \qquad 
    \vec{\pi} \to \vec{\pi} + \Vec{\alpha}_A \sigma.
    \label{6}
\end{equation}
This transformation mixes $\sigma$ and $\vec{\pi}$, and is responsible for the Goldstone boson character of the pions after the spontaneous breaking of chiral symmetry. The Lagrangian density of Eq.~(\ref{eq:1}) is symmetric under the transformations of the chiral group, and from it, one can identify the potential as
\begin{equation}    
    V(\Phi) = -\frac{a^2}{2}(\sigma^2+\vec{\pi}^2) + \frac{\lambda}{4}(\sigma^2+\vec{\pi}^2)^2,
    \label{pot}
\end{equation}
which depends only on the $\text{SU}(2)_L \times \text{SU}(2)_R$ invariant \(\frac{1}{2}\mbox{Tr}\left[\Phi^\dagger\Phi\right] = \sigma^2+\vec{\pi}^2\), and is therefore fully symmetric. Nevertheless, the state \(\langle\sigma\rangle=\langle\vec{\pi}\rangle=0\) is unstable, and the spontaneous symmetry breaking occurs when the vacuum acquires a nonzero expectation value for one of the components $\langle \Phi \rangle = v$, and one customarily sets this direction as $\langle \sigma \rangle = v,\; \langle \vec{\pi} \rangle = 0$, which preserves a $\text{SU}(2)_V$ symmetry subgroup corresponding to isospin rotations of the pions.

Shifting the sigma field, \(\sigma \to \sigma + v\), we obtain the Lagrangian density in the broken phase. From Goldstone's theorem, the three pions emerge as the Goldstone bosons of the spontaneously broken symmetry. To give mass to the pions, we add an explicit symmetry-breaking term \(-h(\sigma+v)\), with $h=m_0^2f_\pi$ and $m_0$ the physical pion mass in vacuum, accounting for the partial conservation of axial current (PCAC) relation. A quark chemical potential \(\mu\) is introduced via the covariant derivative \(\partial_{\mu}\psi \to \partial_{\mu}\psi - i\mu\delta_{\mu}^{0}\psi\). After these modifications, the Lagrangian density becomes
\begin{equation}
    \begin{aligned}
        \lag =& \frac{1}{2}\big[(\partial_{\mu}\sigma)^2+(\partial_{\mu}\vec{\pi})^2\big] - \frac{m_{\sigma}^{2}}{2}\sigma^{2} - \frac{m_{\pi}^{2}}{2}\vec{\pi}^{2} \\
        &+ \bar{\psi}\big(i\slashed{\partial}-m_f\big)\psi + \mu\bar{\psi}\gamma^{0}\psi + \lag_{\text{int}} - V_{\text{tree}},
    \end{aligned}   
    \label{Eq: Lag complete}
\end{equation}
where
\begin{eqnarray}
m_\sigma^2&=&3\lambda v^2 - a^2,\nonumber\\
m_\pi^2&=&\lambda v^2 - a^2,\nonumber\\
m_f&=&gv,
\label{masses}
\end{eqnarray}
are the tree-level masses, and the tree-level potential is given by
\begin{equation}    
    V_{\text{tree}} = -\frac{a^{2}}{2}v^{2} + \frac{\lambda}{4}v^{4} - hv,
    \label{Eq: tree level}
\end{equation}
which is independent of \(\mu\); the dependence on the quark chemical potential enters through loop corrections. The interaction Lagrangian is
\begin{equation}
    \begin{aligned}
        \lag_{\text{int}} =&-\frac{\lambda}{4}(\sigma^2 + \vec{\pi}^2)^2 - \lambda v (\sigma^3 + \sigma \vec{\pi}^2)\\
        &- ig \bar{\psi} \gamma^5 \vec{\tau} \cdot \vec{\pi} \psi - g \bar{\psi} \psi \sigma.
    \end{aligned}
    \label{Eq: Lag int}
\end{equation}
\subsection{Dispersion relations in the presence of \texorpdfstring{$\mu$}{\textmu}}

Since the meson sector is not affected by the finite chemical potential, the dispersion relation takes on the usual form for a relativistic excitation, namely
\begin{equation}
    E_{\sigma} = \sqrt{\mathbf{k}^{2}+m_{\sigma}^{2}}, \qquad 
    E_{\pi} = \sqrt{\mathbf{k}^{2}+m_{\pi}^{2}} .
    \label{Eq: 12}
\end{equation}

\begin{equation}
    D^{-1}_{b}(k) = k^2 + m_b^2.
    \label{Eq: 13}
\end{equation}

For the fermions, the quadratic part of the Lagrangian implies that the inverse fermion propagator in momentum space is given by
\begin{equation}
    S_{f}^{-1}(k) = (k_{0}+\mu)\gamma^{0} - \mathbf{k}\cdot\boldsymbol{\gamma} - m_f .    
    \label{Eq: 14}
\end{equation}
Taking the determinant in the Dirac basis yields 
\begin{equation}
    \det\big[S_{f}^{-1}(k)\big] = \left[m_f^{2} + \mathbf{k}^{2} - (k_{0}+\mu)^{2}\right]^2,
\end{equation}
and setting the determinant to zero determines the fermion dispersion relation to be
\begin{equation}
    \widetilde{E}_{f} = \sqrt{\mathbf{k}^{2}+m_f^{2}} - \mu \equiv E_f - \mu .
    \label{Eq: 15}
\end{equation}

\subsection{One-loop potential}

The one-loop effective potential is expressed in terms of the inverse propagators as
\begin{equation}
    \begin{aligned}
        V_{f}^{(1)} &= i N_f N_c \Omega^{-1} \ln\det\big(S_{f}^{-1}\big), \\[2mm]
        V_{b}^{(1)} &= -\frac{i}{2} \Omega^{-1}\Big[3\ln\det\big(D_{\pi}^{-1}\big) + \ln\det\big(D_{\sigma}^{-1}\big)\Big].
    \end{aligned}
\end{equation}
where $\Omega = \int d^4x$ is the space-time volume. The factors account for fermion degrees of freedom (\(N_f=2\), \(N_c=3\)), the three pions and the sigma. Writing each determinant as an integral containing the corresponding dispersion relation, we obtain
\begin{equation}
\begin{aligned}
V_{f}^{(1)} &= 2i N_f N_c \int\frac{d^{4}k}{(2\pi)^{4}} \ln\!\left[(k_{0}-\mu)^{2} - E_f^{2}\right], \\[2mm]
V_{b}^{(1)} &= -\frac{i}{2}\int\frac{d^{4}k}{(2\pi)^{4}}\Big[3\ln\!\big(k_{0}^{2}-E_{\pi}^{2}\big) + \ln\!\big(k_{0}^{2}-E_{\sigma}^{2}\big)\Big].
\end{aligned}
\end{equation}
The integrals are evaluated using the Matsubara formalism, further detailed in Sec.~\ref{sec:matsubara}. After summing over the fermion Matsubara frequencies \(\tilde{\omega}_n = (2n+1)\pi T \), we get
\begin{eqnarray}
        V_{f}^{(1)} &= &2 N_f N_c \int\frac{d^{3}k}{(2\pi)^{3}} \left\{\frac{}{}E_f
        - T  \ln\!\left[1+e^{-\beta(E_f-\mu)}\right]\right.\nonumber\\
        &-&\left. T \ln\!\left[1+e^{-\beta(E_f+\mu)}\right]\frac{}{}\right\},
    \label{eq:15}
\end{eqnarray}
where $\beta = 1/T$. The first term in Eq.~(\ref{eq:15}) is the vacuum (zero-point) energy, whereas the remaining two terms correspond to the matter contributions. The vacuum term is ultraviolet divergent and must be regularized. Using dimensional regularization in the $\overline{\text{MS}}$ scheme, one finds
\begin{eqnarray}
        V_{vac,f}^{(1)} &= &2N_f N_c \int\frac{d^{3}k}{(2\pi)^{3}} \sqrt{\mathbf{k}^{2}+m_f^{2}}\nonumber\\ 
        &=& -N_f N_c\frac{m_f^{4}}{16\pi^{2}}\left[\frac{1}{\epsilon} + \frac{3}{2} + \ln\left(\frac{\Lambda^{2}}{m_f^{2}}\right)\right],
    \label{Eq: vac fermions}
\end{eqnarray}
where $\Lambda$ is the dimensional regularization ultraviolet scale. Proceeding in a similar fashion for the bosons, for which the Matsubara frequencies are \(\omega_n = 2n\pi T\), we obtain 
\begin{equation}
    V_{b}^{(1)} = -\frac{1}{2}\sum_{b=\sigma,\vec{\pi}}\int\frac{d^{3}k}{(2\pi)^{3}}\Big[ E_b - 2T\ln\!\big(1-e^{-\beta E_b}\big) \Big].
\end{equation}
Here, the first term corresponds to the vacuum contribution, whereas the second term represents the matter contribution. As for the fermion case, we can compute the ultraviolet divergent vacuum piece of the boson contribution which, in dimensional regularization and in the $\overline{\text{MS}}$ scheme, is given by
\begin{equation}
    V_{vac,b}^{(1)} = \sum_{b=\sigma,\vec{\pi}}\frac{m_b^{4}}{64\pi^2}\left[\frac{1}{\epsilon} + \frac{3}{2} + \ln\left(\frac{\Lambda^{2}}{m_b^{2}}\right)\right].
    \label{Eq: vac boson}
\end{equation}
Notice that after renormalization, the overall vacuum contribution,  Eqs.~(\ref{Eq: vac fermions}) and~(\ref{Eq: vac boson}), should not affect either the position or the curvature of the effective potential at the minimum. Their dependence on the condensate $v$ merely renormalizes the tree-level parameters $a^2$ and $\lambda$. Consequently, the vacuum part does not influence the determination of the equilibrium value of $v$ at finite $\mu$ and $T$. The relevant physics for chiral restoration in a medium is encoded entirely in the \emph{matter term}, which is finite and depends explicitly on $\mu$. This term drives the shift of the minimum as the chemical potential increases, eventually leading to the restoration of chiral symmetry.

\subsection{Resummation}

In the medium, the vacuum propagators, Eqs.~(\ref{Eq: 13}) -- (\ref{Eq: 14}), are modified. The modifications are due to interactions which are encoded in self-energies. Let $G$ and $G_0$ be the dressed and free propagators, respectively for a given particle. Generically, the Dyson series relates these propagators by 
\begin{eqnarray}
        G(k) &=&
        \frac{i}{G_0^{-1} - \Pi(k)},
\end{eqnarray}
where $\Pi$ is the corresponding self-energy. Changes in one of the particle's dispersion relation affect the rest of them, since their self-energies depend on each other. The self-consistent solution of the system of coupled equations is tantamount to the resummation procedure. However, in this case, such resummation goes beyond the static ring diagram corrections, since diagrams with an external momentum dependence (see~Fig.\ref{fig1}) contribute to the particle's mass modifications. We now proceed to show how this approach is implemented for the $T=0,\ \mu \neq 0$ case.

\subsection{Zero-temperature limit}

In the limit \(T\to 0\) (\(\beta\to\infty\)) the thermal logarithms simplify such that
\begin{equation}
    \begin{aligned}
        \lim_{\beta\to\infty} T\ln\!\big(1+e^{-\beta(E_f-\mu)}\big) &\to - (E_f-\mu)\,\Theta(\mu-E_f), \\
        \lim_{\beta\to\infty} T\ln\!\big(1+e^{-\beta(E_f+\mu)}\big) &\to 0, \\
        \lim_{\beta\to\infty} T\ln\!\big(1-e^{-\beta E_b}\big) &\to 0.
    \end{aligned}
    \label{simplify}
\end{equation}
Thus, only the fermion part of the matter contribution survives when $\mu > E_f$, and it is the condition that restricts the momentum integration to the Fermi momentum $k_F = \sqrt{\mu^{2}-m_f^{2}}$. The zero-temperature effective potential becomes
\begin{eqnarray}
    V^{(1)} & = &V_f^{(1)} + V_b^{(1)} \nonumber\\
&=& - N_f N_c \frac{\Theta(\mu-M_f)}{\pi^{2}} \nonumber\\
& \times &\int_{0}^{\sqrt{\mu^{2}-M_f^{2}}} dk\,k^{2}\,\big(\mu-\sqrt{k^{2}+M_f^{2}}\big).
\end{eqnarray}
Notice that in the above integral, for the fermion mass we have written $M_f$, to emphasize that as $\mu$ changes, the fermion mass becomes dynamical, that is, $\mu$-dependent. The full effective potential up to one-loop level is given by 


\begin{eqnarray}
       V_{\text{eff}} &=& -\frac{a^{2}}{2}v^{2}+\frac{\lambda}{4}v^{4}-hv - N_f N_c \frac{\Theta(\mu-M_f)}{\pi^{2}}\nonumber\\
        & \times& \int_{0}^{\sqrt{\mu^{2}-M_f^{2}}} dk\,k^{2}\,\big(\mu-\sqrt{k^{2}+M_f^{2}}\big).
        \label{oneloopeffpot}
\end{eqnarray}


\section{One-loop self-energies}\label{III}

\begin{table}[t]
    \centering
    \begin{tabular}{||c|c||}
        \hline
        \multicolumn{2}{|c|}{Vertices of the LSMq} \\
        \hline
        $\sigma^4$ & $-i\lambda/4$ \\
        $\pi^2\sigma^2$ & $-i\lambda/2$ \\
        $\pi^4$ & $-i\lambda/4$ \\
        $\sigma^3$ & $-i\lambda v$ \\
        $\pi^2\sigma$ & $-i\lambda v$ \\
        $\bar{\psi}\pi_i\psi$ & $g\gamma^5\tau_i$ \\
        $\bar{\psi}\sigma\psi$ & $-ig$ \\
        \hline
    \end{tabular}
    \caption{Vertex factors for the particle interactions in the LSMq 
    .}
    \label{tab:vertices}
\end{table}

As the chiral condensate $v$ decreases with increasing quark chemical potential $\mu$, the tree-level masses obtained from Eq.~(\ref{masses}) may become imaginary, signaling the breakdown of the classical description. To cure this issue and obtain physically meaningful masses in the vicinity of the chiral transition, we must include the one-loop corrections to the propagators. These corrections, obtained from the corresponding particle self-energies, incorporate the effects of quantum fluctuations and the surrounding medium into the effective masses. In this section we compute the one-loop self-energies for the sigma, the pion, and the quark using the Matsubara formalism at finite temperature $T$ and quark chemical potential $\mu$. The zero-temperature limit is taken at the end.

\subsection{Interaction vertices and Feynman rules}

From the Lagrangian in the broken phase, the interaction terms that contribute to the self-energies at one loop, Eq.~(\ref{Eq: Lag int}), can be identified. These interactions can be classified into three categories:
\begin{itemize}
    \item \textbf{Four-meson interactions}: $\sigma^4$, $\sigma^2\pi^2$, $\pi^4$ (quartic couplings).
    \item \textbf{Three-meson interactions}: $\sigma^3$, $\sigma\pi^2$ (cubic couplings arising from spontaneous symmetry breaking).
    \item \textbf{Fermion-meson interactions}: $\bar{\psi}\sigma\psi$, $\bar{\psi}\pi_i\psi$ (Yukawa couplings).
\end{itemize}

The quadratic part of the Lagrangian, Eq.~(\ref{Eq: Lag complete}), originates the propagators. In Minkowski space these are
\begin{subequations}
    \begin{align}
        \Delta_\pi(p_0,\mathbf{p}) &= \frac{i}{p^2 - m_\pi^2 + i\epsilon},\\
        \Delta_\sigma(p_0,\mathbf{p}) &= \frac{i}{p^2 - m_\sigma^2 + i\epsilon},\\  
        \tilde{\Delta}_f(p_0,\mathbf{p}) &= \frac{i}{\gamma^\mu \tilde{p}_\mu - m_f + i\epsilon},
    \end{align}
\end{subequations}
where $\tilde{p}^\mu = (p_0 + \mu, \mathbf{p})$. The  vertex factors for each of the particle interactions are summarized in Table~\ref{tab:vertices}. The self-energy for a given particle corresponds to the amputated one-particle irreducible (1PI) two-point function. 
\begin{figure}[t]
	 \begin{tikzpicture}
	    \begin{feynman}[small]
	    \vertex (i1);
	    \vertex [right=of i1] (a);
	    \vertex [right=of a ] (b);
	    \vertex [right=of b ] (f1);
	    \diagram* {
	        (i1) -- [scalar] (a) -- [fermion, half left](b) -- [scalar] (f1),
	        (a) -- [anti fermion, half right] (b),
	        };
	    \end{feynman}
	\end{tikzpicture} \\
	  \begin{tikzpicture}
	    \begin{feynman}[small]
	    \vertex (i1);
	    \vertex [right=of i1] (a);
	    \vertex [right=of a ] (b);
	    \vertex [right=of b ] (f1);
	    \diagram* {
	        (i1) --  (a) -- [fermion, half left](b) -- (f1),
	        (a) -- [anti fermion, half right] (b),
	        };
	    \end{feynman}
	\end{tikzpicture}\\
	\begin{tikzpicture}
	    \begin{feynman}[small]
	    \vertex (i1);
	    \vertex [right=of i1] (a);
	    \vertex [right=of a ] (b);
	    \vertex [right=of b ] (f1);
	    \diagram* {
	        (i1) -- [fermion](a) -- [fermion] (b) -- [fermion] (f1),
	        (a) -- [scalar, half left] (b),
	        };
	    \end{feynman}
	\end{tikzpicture}\, +\, \begin{tikzpicture}
	    \begin{feynman}[small]
	    \vertex (i1);
	    \vertex [right=of i1] (a);
	    \vertex [right=of a ] (b);
	    \vertex [right=of b ] (f1);
	    \diagram* {
	        (i1) -- [fermion](a) -- [fermion] (b) -- [fermion] (f1),
	        (a) -- [ half left] (b),
	        };
	    \end{feynman}
	\end{tikzpicture}
    \caption{Pion (dashed line), sigma (solid line), and quark (solid line with an arrow) one-loop Feynman diagrams contributing to these particles mass modifications. Only 1PI diagrams with quark propagators are non-vanishing at $T=0$ and finite $\mu$.} 
    \label{fig1}
\end{figure}
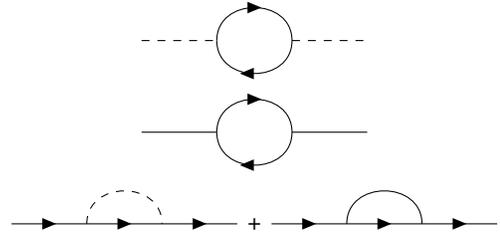
\subsection{Self-energies in Minkowski space}

For each particle, the only non-vanishing one-loop self-energies at $T=0$ and finite $\mu$ are represented by the Feynman diagrams in Fig.~\ref{fig1},     
and their explicit expressions are given by
\begin{subequations}
\begin{eqnarray}
        -i\Pi_{\pi} &=& 4N_c N_f g^2 \int \frac{d^4q}{(2\pi)^4} \frac{1}{\Tilde{q}^2 - m_f^2} \nonumber\\
        &\times&\left[1 + \frac{\Tilde{q}_\mu k^\mu  - k^2}{(k-\Tilde{q})^2 - m_f^2} \right],
    \label{Eq: SE pion}
\end{eqnarray}
\begin{eqnarray}
        -i\Pi_\sigma &=& 4N_c N_f g^2 \int \frac{d^4q}{(2\pi)^4} \frac{1}{\Tilde{q}^2 - m_f^2} \nonumber \\
        &\times&\left[1 + \frac{\Tilde{q}_\mu k^\mu  - k^2}{(k-\Tilde{q})^2 - m_f^2} \right],
    \label{Eq: SE sigma}
\end{eqnarray}
\begin{eqnarray}
        -i\Pi_\Psi &=& -3g^2 \int\frac{d^4q}{(2\pi)^4}\frac{q_0}{\left[(k-\tilde{q})^2 - m_f^2\right](q^2 - m_\pi^2)} \nonumber\\
        &-& g^2 \int\frac{d^4q}{(2\pi)^4}\frac{q_0-2m_f}{\left[(k-\tilde{q})^2 - m_f^2\right](q^2 - m_\sigma^2)},
    \label{Eq: SE fermion}
\end{eqnarray}
    \label{Eq: SE euclides}
\end{subequations}
where $k=(k_0,\mathbf{k})$ is the four-momentum feeding into the loop. We now proceed to the explicit computation of the above expressions at finite temperature and baryon density. Since we are looking for the particle pole masses, we need to set $\mathbf{k}=0$ and $k_0 = M_i$, where $i$ stands for $\sigma$, $\pi$ or the quark

\subsection{Matsubara formalism}\label{sec:matsubara}

In the imaginary-time formalism of finite temperature field theory, the time $\tau$ is compactified to the interval $[0,\beta]$, with $T=1/\beta$. Fields satisfy periodic (bosons) or anti-periodic (fermions) boundary conditions
\begin{equation}
\phi(\tau) = \phi(\tau+\beta), \qquad \psi(\tau) = -\psi(\tau+\beta).
\label{boundary}
\end{equation}
Consequently, the zeroth component of the momentum becomes discrete
\begin{equation}
p_0 \to i\omega_n,\qquad \omega_n = 
\begin{cases}
2n\pi T & \text{(bosons)},\\
(2n+1)\pi T & \text{(fermions)}.
\end{cases}
\label{continuation}
\end{equation}
The loop integrals are then replaced by Matsubara sums:
\begin{equation}
    \int\frac{d^4p}{(2\pi)^4} \to iT\sum_n\int\frac{d^3p}{(2\pi)^3} ,
\label{inttosum}
\end{equation}
with $n$ spanning all integers. Applying this prescription to Eqs.~(\ref{Eq: SE euclides}), we obtain the self-energies in terms of Matsubara sums. It is convenient to introduce the following notation for the propagators
\begin{subequations}
    \begin{equation}
        \Delta_b(\omega_n,E_b) = \frac{1}{\omega_n^2 + E_b^2},
    \end{equation}
    \begin{equation}
        \begin{aligned}
            \tilde{\Delta}(i\tilde{\omega}_n +\mu,E_f) =& \frac{1}{(\tilde{\omega}_n - i\mu)^2 + E^2}.
        \end{aligned}
    \end{equation}
\end{subequations}

Using these expressions, the self-energies become
\begin{widetext}
    
\begin{subequations}
    \begin{align}
        &\begin{aligned}[b]
            \Pi_{\pi} =& 4 N_c N_f g^2 T \sum_n \int \frac{d^3p}{(2\pi)^3} \Tilde{\Delta} (i\Tilde{\omega}_n+\mu,E_f)\\
            -&4 N_c N_fg^2 m_\pi T \sum_n \int \frac{d^3p}{(2\pi)^3} (i\Tilde{\omega}_n +\mu)\Tilde{\Delta}(i\Tilde{\omega}_n+\mu,E_f)\Tilde{\Delta}(i(\omega_m-\Tilde{\omega}_n)-\mu,E_f)\\
            -&4 N_c N_fg^2(2 m_f^2-m_\pi^2)T \sum_n \int \frac{d^3p}{(2\pi)^3} \Tilde{\Delta}(i\Tilde{\omega}_n+\mu,E_f)\Tilde{\Delta}(i(\omega_m-\Tilde{\omega}_n)-\mu,E_f),
        \end{aligned}\\
        &\begin{aligned}[b]
            \Pi_{\sigma} &= 4 N_c N_f g^2 T \sum_n \int \frac{d^3p}{(2\pi)^3} \Tilde{\Delta}(i\Tilde{\omega}_n + \mu ,E_f)\\
            &-4 N_c N_f g^2m_\sigma  T \sum_n \int \frac{d^3p}{(2\pi)^3}  (i\Tilde{\omega}_n +\mu)\Tilde{\Delta}(i\Tilde{\omega}_n+\mu,E_f)\Tilde{\Delta}(i(\omega_m-\Tilde{\omega}_n)-\mu,E_f)\\
            &+4 N_c N_f g^2m_\sigma^2  T \sum_n \int \frac{d^3p}{(2\pi)^3}  \Tilde{\Delta}(i\Tilde{\omega}_n+\mu,E_f)\Tilde{\Delta}(i(\omega_m-\Tilde{\omega}_n)-\mu,E_f),
        \end{aligned}\\
        &\begin{aligned}[b]
            \Pi_{\Psi} =& g^2 T \sum_n \int \frac{d^3p}{(2\pi)^3}i\omega_n\Tilde{\Delta}(i(\Tilde{\omega}_m-\omega_n)+\mu, E_f)\left\{ 3\Delta(i\omega_n, E_\pi) + \Delta(i\omega_n, E_\sigma)\right\}\\
            -&2m_f g^2  T \sum_n \int \frac{d^3p}{(2\pi)^3} \Delta(i\omega_n, E_\sigma)\Tilde{\Delta}(i(\Tilde{\omega}_m-\omega_n)+\mu, E_f).
        \end{aligned}
    \end{align}
    \label{Eq: 30}
\end{subequations}

\subsection{Summation over Matsubara frequencies}

The sums over Matsubara frequencies can be performed analytically using standard techniques. The key identities are
\begin{subequations}
    \begin{align}
        T \sum_n \Tilde{\Delta}(i\Tilde{\omega}_n + \mu ) &= \frac{1}{2E} \left[ 1 -\Tilde{f}(E + \mu) -\Tilde{f}(E-\mu) \right],\\
        T\sum_n \Tilde{\Delta}(i\Tilde{\omega}_n +\mu,E_f)\Tilde{\Delta}(i(\omega_m-\Tilde{\omega}_n)-\mu,E_f) &=  -\sum_{s_1,s_2 = \pm}\frac{s_1s_2}{4E_f^2} \frac{ 1-\Tilde{f}(s_1E_f -\mu)-\Tilde{f}(s_2E_f +\mu)}{i\omega_m-s_1E_f-s_2E_f},\\
        T\sum_n (i\Tilde{\omega}_n+\mu)\Tilde{\Delta}(i\Tilde{\omega}_n +\mu,E_f)\Tilde{\Delta}(i(\omega_m-\Tilde{\omega}_n)-\mu,E_f) &=  -\sum_{s_1,s_2 = \pm}\frac{ s_2}{4E_f}\frac{\left[1-\Tilde{f}(s_1E_f -\mu)-\Tilde{f}(s_2E_f +\mu)\right]}{i\omega_m-s_1E_f-s_2E_f},\\
        T\sum_n \Delta(i\omega_n,E_b)\Tilde{\Delta}(i(\Tilde{\omega}_m-\omega_n)+\mu,E_f) &=  - \sum_{s_1,s_2 = \pm}\frac{s_1s_2}{4E_bE_f}\frac{\left[1+f(s_1E_b)-\Tilde{f}(s_2E_f -\mu)\right]}{i\Tilde{\omega}_m-s_1E_b-s_2E_f+\mu},\\
        T\sum_n (i\omega_n)\Delta(i\omega_n,E_b)\Tilde{\Delta}(i(\Tilde{\omega}_m-\omega_n)+\mu,E_f) &=   -\sum_{s_1,s_2 = \pm}\frac{s_2}{4E_f}\frac{\left[1+f(s_1E_b)-\Tilde{f}(s_2E_f -\mu)\right]}{i\omega_m-s_1E_b-s_2E_f+\mu},
		\end{align}
    \label{identities}
\end{subequations}
where $f(E) = 1/(e^{\beta E}-1)$ is the Bose-Einstein distribution and $\tilde{f}(E) = 1/(e^{\beta E}+1)$ is the Fermi-Dirac distribution. Applying these identities to Eqs.~(\ref{Eq: 30}) we obtain the self-energies in their final form before renormalization
\begin{subequations}
\begin{eqnarray}
\Pi_\pi &=&\; 4N_cN_f g^2 \int\frac{d^3q}{(2\pi)^3}\frac{1}{2E_f}\Big[1 - \tilde{f}(E_f+\mu) - \tilde{f}(E_f-\mu)\Big] \nonumber\\
&+& 4N_cN_f g^2 m_\pi \int\frac{d^3q}{(2\pi)^3}\frac{1 - \tilde{f}(E_f-\mu) - \tilde{f}(E_f+\mu)}{4E_f}\left(\frac{1}{i\omega_m-2E_f} + \frac{1}{i\omega_m+2E_f}\right) \nonumber\\
&-& 4 N_c N_fg^2 (m_\pi^2-2m_f^2) \int\frac{d^3q}{(2\pi)^3}\frac{1 - \tilde{f}(E_f-\mu) - \tilde{f}(E_f+\mu)}{4E_f^2}\left(\frac{1}{i\omega_m-2E_f} - \frac{1}{i\omega_m+2E_f}\right),
\label{selfpiexpl}
\end{eqnarray}
\begin{eqnarray}
\Pi_\sigma &=&\; 4N_cN_f g^2 \int\frac{d^3q}{(2\pi)^3}\frac{1}{2E_f}\Big[1 - \tilde{f}(E_f+\mu) - \tilde{f}(E_f-\mu)\Big] \nonumber\\
&+& 4N_cN_f g^2 m_\sigma \int\frac{d^3q}{(2\pi)^3}\frac{1 - \tilde{f}(E_f-\mu) - \tilde{f}(E_f+\mu)}{4E_f}\left(\frac{1}{i\omega_m-2E_f} + \frac{1}{i\omega_m+2E_f}\right)\nonumber\\
&-& 4 N_c N_f g^2m_\sigma^2 \int \frac{d^3p}{(2\pi)^3}\frac{1-\Tilde{f}(E_f-\mu)-\Tilde{f}(E_f+\mu)}{4E_f^2} \left( \frac{1}{i\omega_m-2E_f} - \frac{1}{i\omega_m+2E_f}\right),
\label{selfsigmaexpl}
\end{eqnarray}
\begin{eqnarray}
\Pi_\Psi &=&\; -3g^2 \int\frac{d^3q}{(2\pi)^3}\frac{1}{4E_f}\Bigg\{\frac{1+f(E_\pi)-\tilde{f}(E_f-\mu)}{i\omega_m-E_\pi-E_f+\mu} + \frac{1+f(E_\pi)-\tilde{f}(E_f+\mu)}{i\omega_m+E_\pi+E_f+\mu}\Bigg\} \nonumber\\
&+& 3g^2 \int\frac{d^3q}{(2\pi)^3}\frac{1}{4E_f}\Bigg\{\frac{f(E_\pi)+\tilde{f}(E_f+\mu)}{i\omega_m-E_\pi+E_f+\mu} + \frac{f(E_\pi)+\tilde{f}(E_f-\mu)}{i\omega_m+E_\pi-E_f+\mu}\Bigg\} \nonumber\\
&-& g^2 \int\frac{d^3q}{(2\pi)^3}\frac{1}{4E_f}\Bigg\{\frac{1+f(E_\sigma)-\tilde{f}(E_f-\mu)}{i\omega_m-E_\sigma-E_f+\mu} + \frac{1+f(E_\sigma)-\tilde{f}(E_f+\mu)}{i\omega_m+E_\sigma+E_f+\mu}\Bigg\} \nonumber\\
&+& g^2 \int\frac{d^3q}{(2\pi)^3}\frac{1}{4E_f}\Bigg\{\frac{f(E_\sigma)+\tilde{f}(E_f+\mu)}{i\omega_m-E_\sigma+E_f+\mu} + \frac{f(E_\sigma)+\tilde{f}(E_f-\mu)}{i\omega_m+E_\sigma-E_f+\mu}\Bigg\} \nonumber\\
&-& 2m_f g^2 \int\frac{d^3q}{(2\pi)^3}\frac{1}{4E_\sigma E_f}\Bigg\{\frac{1+f(E_\sigma)-\tilde{f}(E_f-\mu)}{i\tilde{\omega}_m-E_\sigma-E_f+\mu} - \frac{1+f(E_\sigma)-\tilde{f}(E_f+\mu)}{i\tilde{\omega}_m+E_\sigma-E_f+\mu}\Bigg\} \nonumber\\
&-& 2m_f g^2 \int\frac{d^3q}{(2\pi)^3}\frac{1}{4E_\sigma E_f}\Bigg\{\frac{f(E_\sigma)+\tilde{f}(E_f+\mu)}{i\tilde{\omega}_m-E_\sigma+E_f+\mu} - \frac{f(E_\sigma)+\tilde{f}(E_f-\mu)}{i\tilde{\omega}_m+E_\sigma-E_f+\mu}\Bigg\}.
\label{selfferexpl}
\end{eqnarray}
\end{subequations}
\subsection{Zero-temperature limit}

In the $T \to 0$ limit, the distribution functions simplify as in Eq.~(\ref{simplify}). The Bose-Einstein distribution functions vanish, while the Fermi-Dirac distribution functions become step functions:
\begin{eqnarray}
\lim_{T\to 0} \tilde{f}(E_f-\mu) &=& \Theta(\mu - E_f), \nonumber\\
\lim_{T\to 0} \tilde{f}(E_f+\mu) &=& 0, \nonumber\\
\lim_{T\to 0} f(E_b) &=& 0.
\end{eqnarray}
Substituting these limits into Eqs.~(\ref{selfferexpl}), we notice that the first terms in each of these equations become proportional to the vacuum integrals and thus can be absorbed into the renormalization of the masses. The remaining terms are finite and represent the medium corrections. To compute the pole masses, we take the real part of the above self-energies and set the zeroth component of the external momentum equal to the corresponding pole mass. After a straight forward computation, one gets the final form for the dynamical masses
\begin{subequations}
\begin{equation}
M_\pi^2 = (\lambda v^2 - a^2) + \frac{4N_cN_f g^2}{\pi^2} \int_0^{\sqrt{\mu^2 - M_f^2}} dq\,\frac{q^4}{\sqrt{q^2 + M_f^2}}\,\frac{1}{M_\pi^2 - 4(q^2 + M_f^2)},
\end{equation}
\begin{equation}
M_\sigma^2 = (3\lambda v^2 - a^2) + \frac{4N_cN_f g^2}{\pi^2} \int_0^{\sqrt{\mu^2 - M_f^2}} dq\,\frac{q^2\sqrt{q^2 + M_f^2}}{M_\sigma^2 - 4(q^2 + M_f^2)},
\label{masssigma}
\end{equation}
\begin{eqnarray}
M_f &=& gv + \frac{3g^2}{4\pi^2} \int_0^{\sqrt{\mu^2 - M_f^2}} dp\,\frac{p^2}{\sqrt{p^2 + M_f^2}}\,\frac{M_f - \sqrt{p^2 + M_f^2} + \mu}{(M_f - \sqrt{p^2 + M_f^2} + \mu)^2 - (p^2 + M_\pi^2)}\nonumber\\
&+& \frac{g^2}{4\pi^2} \int_0^{\sqrt{\mu^2 - M_f^2}} dp\,\frac{p^2}{\sqrt{p^2 + M_f^2}}\,\frac{3M_f - \sqrt{p^2 + M_f^2} + \mu}{(M_f - \sqrt{p^2 + M_f^2} + \mu)^2 - (p^2 + M_\sigma^2)}.
\label{massfer}
\end{eqnarray}
\label{Eq: 34}
\end{subequations}
\end{widetext}
Once again, notice that in the above equations, we have replaced the vacuum masses ($m_\pi,\ m_\sigma,\ m_f$) by medium-modified masses ($M_\pi,\ M_\sigma,\ M_f$) to emphasize that as $\mu$ changes, masses become dynamical, that is, $\mu$-dependent. The set of equations~(\ref{Eq: 34}) must be solved simultaneously, along with the minimization condition
\begin{equation}
\frac{d V_{\text{eff}}}{dv} = 0,
\label{gapeq}
\end{equation}
where $V_{\text{eff}}$ is given by Eq.~(\ref{oneloopeffpot}).
The system of equations (\ref{Eq: 34}) and~(\ref{gapeq}) constitutes the full one-loop self-consistent calculation of the particle masses in the two-flavor version of the LSMq at finite baryon density and zero temperature. The integrals are finite and have support only on the Fermi surface $k < \sqrt{\mu^2 - M_f^2}$, which naturally vanishes when $\mu < M_f$. Solving these equations numerically yields the density-dependent masses and chiral condensate, providing a framework to study chiral symmetry restoration while eliminating the issue of negative squared meson masses when these are treated at tree-level. We now proceed to implement the numerical solution to find the $\mu$ dependence of these masses.

\section{\texorpdfstring{$\mu$}{mu}-dependent particle masses}\label{IV}

\begin{figure}[t]
    \centering
    \includegraphics[width=0.9\linewidth]{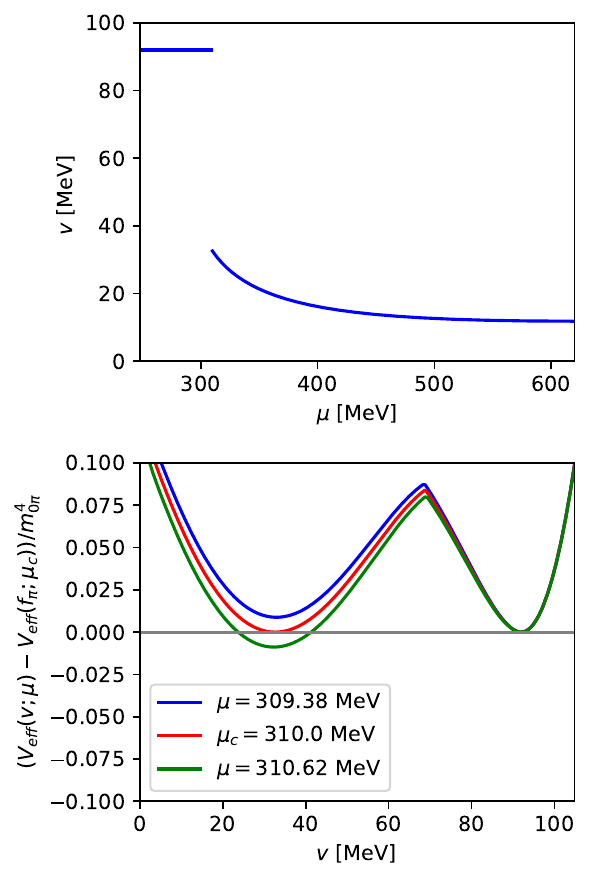}
    \caption{(top) Chiral condensate $v$ as a function of the quark chemical potential $\mu$ at zero temperature. The condensate remains constant at its vacuum value $v = 92$~MeV for $\mu \lesssim m_{0f} = 310$~MeV. At this value the condensate presents a discontinuity after which it decreases monotonically as $\mu$ increases. The decreasing trend continues smoothly, reaching $v \approx 11.84$~MeV at $\mu = 620$~MeV, indicating a gradual restoration of chiral symmetry. (bottom) Effective potential as a function of the order parameter. The minima of the effective potential become degenerate at the critical value of the quark chemical potential $\mu_c$ at which point the order parameter experiences a discontinuous transition. The parameters used in the calculation are $f_\pi = 92$~MeV, $m_{0\pi} = 140$~MeV, $m_{0\sigma} \approx 635.61$~MeV, and $g_0 \approx 3.37$, with the self-consistent masses determined from Eqs.~(\ref{Eq: 34}).}
    \label{fig:chiral cond}
\end{figure}

To find the solution of the system described by Eqs.~(\ref{Eq: 34}) - (\ref{gapeq}) it is necessary to implement a self-consistent numerical approach. The problem can be decomposed into two interconnected parts: (i) for a fixed value of the chiral condensate $v$ and quark chemical potential $\mu$, the dynamical masses $M_f$, $M_\pi$, and $M_\sigma$ are obtained by solving the system of coupled gap equations; (ii) the effective potential $V_{\text{eff}}(v)$ is then evaluated, and the global minimum with respect to $v$ is found. This procedure is repeated for each value of $\mu$ in the considered range.

\subsection{Self-consistent mass determination}

Given that the solutions are more stable for large values of $\mu$, the procedure consists of looking for solutions starting in that regime. Since the search is highly dependent on the initial seed, we use a random generation of guess masses and try to solve the system for a small enough value of $v=0.2f_\pi$. This choice is motivated by the expectation that for large values of $\mu$, chiral symmetry is expected to be restored. The solutions are then filtered to find a more accurate set of solutions. Once this set is found, it is used to solve the system again to find the next best set. This means that we need to use the random seeds generator only once at the beginning of the procedure, and the subsequent solutions can be found by implementing small variations of the masses and $v$. 

The integrals involved are one-dimensional. Since we are pursuing the real part of the self-energies, the principal value of the integrals is implemented using a small neighborhood of size $\epsilon_{\text{PV}} = 1 \times 10^{-6}$ around the pole. The iteration continues until the relative change between successive iterations falls below a tolerance of $10^{-6}$ using the \lq\lq quad" routine from \lq\lq scipy.integrate\rq\rq\  for numerical integrations and \lq\lq scipy.optimize\rq\rq\ for the non-linear equation solutions in Python.

\begin{figure}[t]
    \centering
    \includegraphics[width=0.9\linewidth]{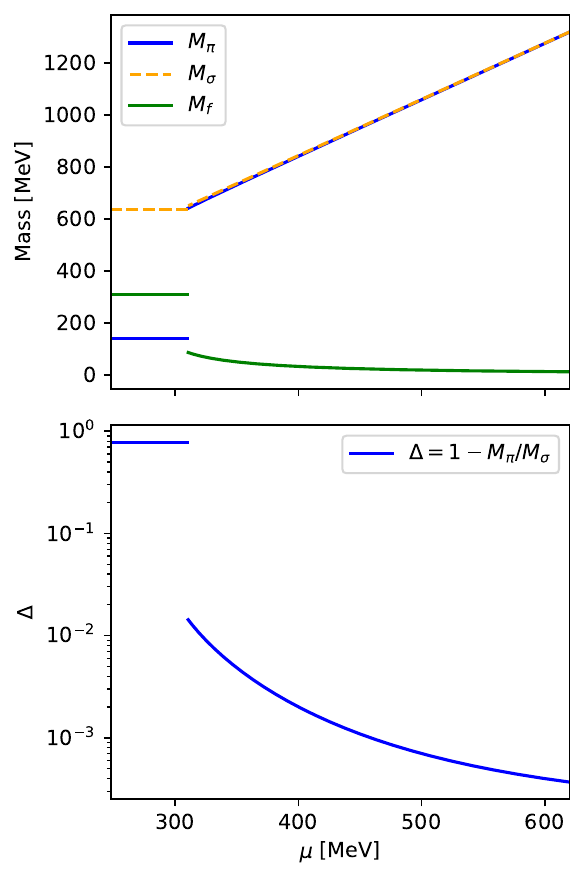}
    \caption{(top) Dynamical masses as functions of the quark chemical potential $\mu$ at zero temperature. The pion, sigma and quark masses $M_\pi$, $M_\sigma$ and $M_f$, respectively, remain constant and at its vacuum values up to $\mu=m_{0f}=310$ MeV at which point they undergo a discontinuous transition. For larger values of $\mu$ the quark mass decreases monotonically. The decrease reflects the melting of the chiral condensate. $M_\pi$ catches up with $M_\sigma$ and both increase with $\mu$ to become degenerate in the high-density limit, signaling the restoration of chiral symmetry. (bottom) The quantity $\Delta = 1 - M_\pi/M_\sigma$ measures the rate of approach to chiral symmetry restoration.}
    \label{fig3}
\end{figure}

\subsection{Minimization of the effective potential}

The minimization of $V_{\text{eff}}(v)$ for a fixed $\mu$ is performed using an adaptive step-size algorithm. The procedure is as follows:

\begin{enumerate}
    \item An initial guess for $v$ is provided for $\mu = \mu_\text{max}$. For subsequent $\mu_i$ values, the previous solution serves as the initial guess.
    \item The self-consistent masses are computed for the current $v$, and $V_{\text{eff}}(v)$ is evaluated.
    \item A step $\epsilon_v = 0.5$~MeV is taken to the left ($v \to v - \epsilon_v$), and the masses and potential are recomputed.
    \item If the new potential is lower than the previous one, the step is repeated in the same direction.
    \item If the new potential is larger, the step size is halved ($\epsilon_v \to \epsilon_v/2$), and a step to the right is attempted.
    \item This process of alternating directions and reducing step size is repeated $N_{\text{recursive}} = 20$ times, yielding a final precision in $v$ of $\epsilon_v / 2^{N_{\text{recursive}}}$.
\end{enumerate}

This algorithm efficiently locates the global minimum of $V_{\text{eff}}(v)$ without requiring explicit computation of the derivative, which is advantageous given the implicit dependence of the masses on $v$.

\subsection{Parameters and numerical implementation}

\begin{figure}[t]
    \centering
    \includegraphics[width=0.9\linewidth]{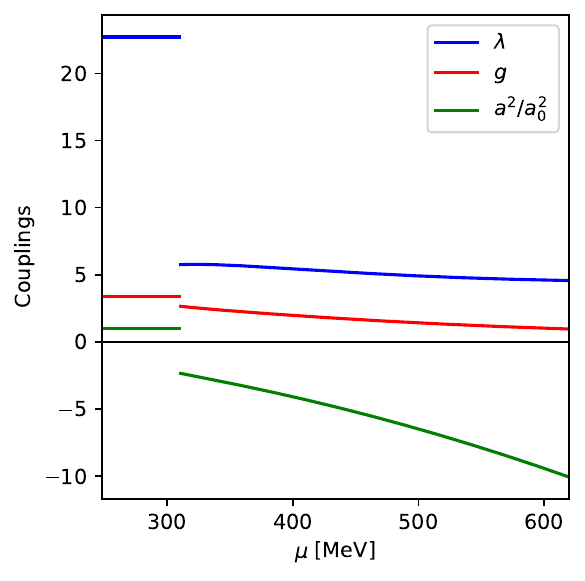}
    \caption{Parameters $\lambda$, $g$, and renormalized $a^2/a_0^2$ with $a_0 = 415.45$~MeV as functions of the quark chemical potential $\mu$ at $T=0$. The vacuum values are $\lambda_0 = 22.71$, $g_0 = 3.37$. These parameters, as well as $m_{0\sigma}$ are fixed once the values of $m_{0\pi}$ and $m_{0f}$ are fixed. The coupling $\lambda$ slightly increases after the phase transition to then monotonically decrease for larger values of $\mu$. The Yukawa coupling $g$ decreases monotonically, reflecting the weakening of the interaction as the chiral symmetry is restored. Most notably, the mass parameter $a^2$ changes sign from positive to negative as $\mu$ increases beyond $310$~MeV. This sign change is a clear signature of chiral symmetry restoration: in the symmetric phase ($\mu > \mu_c$), the Mexican-hat potential reverts to a simple parabolic potential centered at $v$, indicating that the vacuum expectation value $\langle \sigma \rangle$ is unique.}
    \label{fig:Coulplings}
\end{figure}

The model parameters are fixed by the physical conditions in vacuum. The pion decay constant is taken as $f_\pi = 92$~MeV. The constituent quark mass is taken as $m_f = 310$~MeV, and the pion mass as $m_\pi = 140$~MeV. The sigma mass is not independent but satisfies the Ward identity for the partially conserved axial current (PCAC)~\cite{Ayala:2024sqm}
\begin{equation}
    m_\sigma = \sqrt{m_\pi^2 + 4m_f^2} \approx 635.61~\text{MeV}.    
\end{equation}
These values correspond to a Yukawa coupling $g_0 = m_f/f_\pi \approx 3.37$ and a quartic coupling $\lambda_0 = (m_{0\sigma}^2 - m_{0\pi}^2)/(2f_\pi^2) \approx 22.7$. The chemical potential range explored is $\mu \in [0.8\,m_{0f},\; 2\,m_{0f}]$, discretized into 601 equally spaced points. This range covers the transition region and extends into the restored phase.

Figure~\ref{fig:chiral cond} shows the behavior of the chiral condensate $v$ as a function of $\mu$ for $T=0$. The first-order phase transition is signaled by the discontinuous drop of the condensate that occurs for $\mu=m_{0f}$. Figure~\ref{fig3} (top) shows the behavior of the dynamical masses as functions of $\mu$. The first order phase transition is signaled by the discontinuous change in the quark and meson masses. The former decreases, approaching zero for large values of $\mu$. The latter increase and become degenerate for large values of $\mu$. This behavior heralds the restoration of chiral symmetry. The bottom panel in Fig.~\ref{fig3} quantifies the difference between the pion and sigma masses as $\mu$ increases. These masses become almost equal soon after the phase transition. 

Assuming that the functional relation obeyed by the tree-level masses and the couplings, Eqs.~(\ref{masses}), is maintained for the dynamical masses, namely,
\begin{eqnarray}
M_\sigma^2(\mu)&=&3\lambda(\mu) v^2(\mu) - a^2(\mu),\nonumber\\
M_\pi^2(\mu)&=&\lambda(\mu) v^2(\mu) - a^2(\mu),\nonumber\\
M_f(\mu)&=&g(\mu)v(\mu),
\label{dynmassrel}
\end{eqnarray}
one can turn the results for the $\mu$ dependence of the masses into the $\mu$ dependence of the couplings. Figure~\ref{fig:Coulplings} shows the behavior of the couplings $\lambda$ and $g$, as well as the mass parameter $a^2$, as functions of $\mu$. These quantities exhibit a discontinuity at the phase transition.
Both $\lambda$ and $g$ decrease from their vacuum values, with $\lambda$ displaying a more pronounced drop. Most notably, the mass parameter $a^2$ changes sign across the phase transition, going from positive to negative values. Recall that chiral symmetry is realized in the Wigner--Weyl mode when the symmetry of the Lagrangian is also realized in the vacuum state. In the present model, this occurs when the curvature of the effective potential at the origin, controlled by the mass parameter $a^2$ in Eq.~(\ref{pot}), becomes positive. Therefore, the sign change of $a^2$ signals the transition from the spontaneously broken phase to the chirally restored phase.

\subsection{Derived thermodynamic quantities}

From the effective potential, several thermodynamic quantities of interest are derived

\begin{figure}[t]
    \centering
    \includegraphics[width=0.9\linewidth]{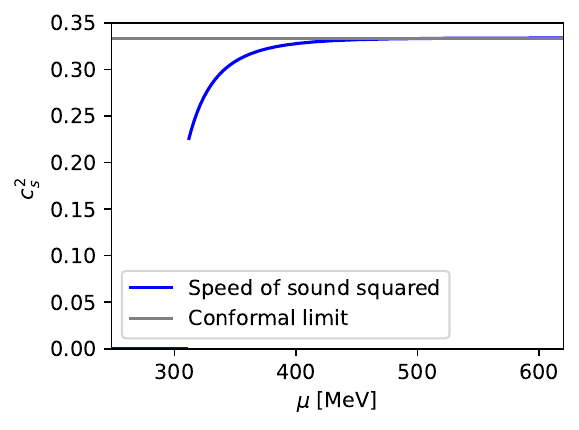}
    \caption{Speed of sound squared $c_s^2 = \partial P/\partial \epsilon$ as a function of the quark chemical potential $\mu$ at $T=0$, compared with the conformal limit $c_s^2 = 1/3$ (continuous horizontal line). At low densities ($\mu \lesssim 310$~MeV), $c_s^2$ is zero, indicating that the medium is made up of hadron matter. As $\mu$ increases beyond $ m_{0f}$, $c_s^2$ rises monotonically and approaches the conformal limit $1/3$ from below at high densities ($\mu \gtrsim 500$~MeV). This behavior suggests that the system undergoes a transition from hadron matter to a regime where the equation of state becomes increasingly conformal, consistent with the restoration of chiral symmetry and the emergence of deconfined quark matter. The approach to $c_s^2 = 1/3$ is a signature of the emergence of a system of non-interacting relativistic massless quarks.}
    \label{fig:speed_of_sound}
\end{figure}

\begin{itemize}
    \item \textbf{Pressure:} $P(\mu) = -V_{\text{eff}}(\mu) + V_{\text{eff}}(\mu=0)$, where the subtraction removes the vacuum energy contribution.
    \item \textbf{Quark number density:} $n_q(\mu) = \dfrac{\partial P}{\partial \mu}$, computed via numerical differentiation.
    \item \textbf{Energy density:} $\epsilon(\mu) = \mu n_q(\mu) - P(\mu)$.
    \item \textbf{Speed of sound squared:} 
    \begin{eqnarray}
    c_s^2(\mu) &=& \dfrac{\partial P}{\partial \epsilon}\nonumber\\
    &=& \dfrac{n_q}{\epsilon + P}\,\dfrac{\partial \mu}{\partial n_q}\nonumber\\
    &=& \dfrac{n_q^2}{(\epsilon + P)\,\partial n_q/\partial \mu}.
    \end{eqnarray}
\end{itemize}
The speed of sound is particularly sensitive to the equation of state and provides a direct probe of the conformal properties of the system. In the limit of a non-interacting gas of massless quarks, one expects $c_s^2 = 1/3$. Figure~(\ref{fig:speed_of_sound}) shows the speed of sound squared as a function of $\mu$. For $\mu\leq m_{0f}$, $c_s^2$ is zero, becoming finite in a discontinuous way at the phase transition and eventually reaching the conformal limit from below for large values of $\mu$.

\section{Summary and conclusions}\label{concl}

In this work, we have used the LSMq based on a SU(2)$_L\times$SU(2)$_R$ chiral symmetry to study the evolution of the quark and meson masses at $T=0$ and finite $\mu$, at one-loop order. To this end, we set up the system of coupled equations satisfied by the particle pole masses and solved them numerically. This procedure requires determining the $\mu$-dependent vacuum expectation value of the scalar sigma field, obtained by minimizing the one-loop effective potential. This approach goes beyond previous treatments, where the self-energy contributions are computed using approximate schemes such as ring-diagram resummation. We find that the phase transition is first order and occurs at $\mu=m_{0f}$, where $m_{0f}$ is the vacuum quark mass. The first-order nature of the phase transition is signaled by the discontinuous behavior of the chiral condensate, particle masses, and couplings. Thermodynamic quantities are obtained from the effective potential. In particular, we have shown that the square of the speed of sound undergoes a discontinuous transition, becoming finite at the phase transition and approaching the conformal limit from below for large values of $\mu$. This procedure lends itself to being extended to explore the effective QCD phase diagram in the $\mu$, $T$ variables, and eventually the effects of other chemical potentials, such as the one associated with an isospin imbalance. This is work that we are currently pursuing and that will soon be reported elsewhere.

\section*{Acknowledgements}
The authors wish to thank Alma T. Aquino for her help implementing the numerical computation. A.A. thanks the colleagues and staff of Universidade de São Paulo, of Instituto de F\'isica Te\'orica, UNESP and of Universidade Cidade de São Paulo for their kind hospitality during a sabbatical stay during which part of this work was carried out. A.A. also acknowledges support from the PASPA program of Direcci\'on General de Asuntos del Personal Acad\'emico (DGAPA) of the Universidad Nacional Aut\'onoma de M\'exico (UNAM) for the sabbatical stay during which this research was carried out and for support via grant number IG100826. Support for this work has been received in part via Secretar\'\i a de Ciencia, Humanidades, Tecnolog\'\i a e Innovaci\'on (SECIHTI) M\'exico grant numbers CIORGANISMOS-2025-17 and CBF-2025-G-1718. R.Z acknowledges support from ANID/CONICYT FONDECYT Regular (Chile) under Grant No. 1241436.
This work was partially supported by Conselho Nacional de Desenvolvimento Cient\'ifico e Tecno\-l\'o\-gico  (CNPq), Grants No. 312032/2023-4, 402963/2024-5 and 445182/2024-5 (R.L.S.F.), No. 141270/2023-3 and 201300/2025-7 (B.~S.~L.), and No.~409032/2023-9 (B.E.); Funda\c{c}\~ao de Amparo \`a Pesquisa do Estado do Rio 
Grande do Sul (FAPERGS), Grant No. 24/2551-0001285-0 (R.L.S.F.); The work is also part of the project
Instituto Nacional de Ci\^encia e Tecnologia -- F\'isica Nuclear e
Aplica\c{c}\~oes (INCT - FNA), Grants No. 464898/2014-5 and 408419/2024-5, and supported
by the Ser\-ra\-pi\-lhei\-ra Institute (grant number Serra -
2211-42230). R. L. S. F. and B.~S.~L. acknowledge the kind hospitality of the Center for Nuclear Research at Kent State University, where part of this work was
done.

\bibliography{biblio}

\end{document}